# Emotion Recognition from Human Speech: Emphasizing on Relevant Feature Selection and Majority Voting Technique

Md. Kamruzzaman Sarker[1], Kazi Md. Rokibul Alam[2] and Md. Arifuzzaman[3]
Department of Computer Science and Engineering
Khulna University of Engineering and Technology
Khulna-9203, Bangladesh
Email: mdkamruzzamansarker@gmail.com[1], rokibcse@yahoo.com[2], arifuzzaman.likhon@gmail.com[3]

*Abstract*— This paper proposes an approach to detect emotion from human speech employing majority voting technique over several machine learning techniques. The contribution of this work is in two folds: firstly it selects those features of speech which is most promising for classification and secondly it uses the majority voting technique that selects the exact class of emotion. Here, majority voting technique has been applied over Neural Network (NN), Decision Tree (DT), Support Vector Machine (SVM) and K-Nearest Neighbor (KNN). Input vector of NN, DT, SVM and KNN consists of various acoustic and prosodic features like Pitch, Mel-Frequency Cepstral coefficients etc. From speech signal many feature have been extracted and only promising features have been selected. To consider a feature as promising, Fast Correlation based feature selection (FCBF) and Fisher score algorithms have been used and only those features are selected which are highly ranked by both of them. The proposed approach has been tested on Berlin dataset of emotional speech [3] and Electromagnetic Articulography (EMA) dataset [4]. The experimental result shows that majority voting technique attains better accuracy over individual machine learning techniques. The employment of the proposed approach can effectively recognize the emotion of human beings in case of social robot, intelligent chat client, call-center of a company etc.

*Index Terms*—Component Emotion recognition, k-nearest neighbor, support vector machine, decision tree, neural network, majority voting technique.

1. INTRODUCTION

With the proliferation of intelligent machine, the requirement of emotion recognition is increasing exponentially. In the area of theoretical science to engineering, emotion recognition has drawn extra concentration. It has been proved that emotion has great impact on decision-making [1] and social communication. Emotion recognition is required in the field of intelligent machine as it can behave as like as human being by taking decisions. Moreover, it can interface with human being for the purpose of eased communication. As emotion is highly important in social communications, it is desirable that intelligent machine would be able to recognize human emotions effectively.

There exist lots of difficulties to detect emotion. Because the no. of basic emotion labels are arguable still now [5, 6] and same emotion can be defined in different ways depending on the situations. Another problem is that the emotion of a particular class may carry the component of other classes. Although there exist numerous complexities, researchers are working to overcome them. Several pattern recognition techniques such as Maximum Likelihood Bayes (MLB) classifier, Kernel Regression (KR), KNN and other methods were used in several times for emotion recognition from speech [8-10]. Many researchers have attempted to solve real time problems also. Moreover, feature set used by researchers was not concrete and even now it varies significantly.

From speech, many acoustic and prosodic features can be obtained. It is possible to detect emotion by using lots of emotional features but recognition time will be higher. So it is necessary to select exact features. In this paper, we have emphasized the feature selection by combining the result of two feature selection algorithms. It gives us 12 best features to classify the class label. By applying best feature set individually on NN, DT, SVM and KNN, accuracy has been obtained up to 81.69%. Then to reduce misclassification rate, majority voting technique has been applied on those techniques to take the final decision, which gives an accuracy of 84%.

2. EXISTING WORKS

The first research on emotion for animal and human being was done by Charles Darwin. After that many researcher have worked out on this topics. Frank Dellaert et al. [8] detected four emotions which were happy, sad, anger and fear from their own dataset. They used 17 selected features from 5 groups and used three methods which were MLB classifier, KR and KNN where the maximum accuracy was found by KNN [8].

Tin Lay New et al. [12] used Hidden Markov model (HMM) to classify 6 categories of emotion and found average accuracy 78%. He used a database consisting of 60 emotional utterances, each from twelve speakers [12]. He used log frequency power co-efficients (LFPC) to represent the speech signals and compared its performance with linear prediction Cepstral coefficients (LPCC) feature parameters and mel-frequency Cepstral coefficients (MFCC) feature parameters.

He found that the better result is obtained by using LFPC features.

Klaus R. Scherer et al. [7] studied a cross-cultural investigation to detect emotion from 9 natural languages and achieved 66% accuracy for five emotional classes which were happy, disgust, sad, anger and fear [7]. He collected dataset from professional Actor and Actress.

Vallery A. Petrushin et al. [9] developed a real time application for call center to detect five emotional classes with accuracy 77% and tested it on Berlin database of emotional speech [3]. For selecting important features he used correlation based feature selection algorithm and selected 7 best features. Feng Yu, et al. [13] applied classification algorithms KNN, NN and SVM on segment of speech from Chinese teleplays, to detect four emotional classes and achieved maximum 74% accuracy using SVM. They used 16 acoustic and prosodic features. Australian researcher Vidhyasaharan Sethu et al. [10] used only acoustic features with warping and without warping technique and HMM to detect five classes of emotion. They achieved mean accuracy of 41.6%.

Björn Schuller et al. [11] tried to rank the acoustic features that contain more information to detect emotion according to a Linear Discriminant Analysis. He showed that most important features are pitch and energy related. They used dataset collected from German and English sentences of 13 actors [11].

Although there exist lots of related works, the major motivation of the proposed approach is to improve accuracy effectively using possibly less no. of features.

## 3. IMPLEMENTATION

Two well-known emotional speech datasets are used here. These are Berlin dataset of emotional speech and EMA dataset. Berlin dataset is classified by 7 emotions: neutral, anger, fear, happiness, sadness, disgust and boredom. From here we have taken 4 emotions: anger, happiness, neutral and sadness. EMA dataset is classified by 4 emotions: neutral, anger, sadness and happiness.

This simulation works in case of predefined test data or live speech. To use live speech, speech signal has to undergo through some preprocessing, which includes segmentation, noise reduction etc. A flow-chart of the proposed approach is shown Figure 1.

### 3.1 Feature Extraction

Many existing works emphasize on fundamental frequency (F0) i.e. pitch to be the main vocal key for emotion. However other keys like vocal energy, frequency spectral feature, speaking rate etc. are also important in the case of emotion recognition.

As initial set of feature, we have extracted 16 low-level descriptors and 1st order delta co-efficient of those descriptors. Then 11 distinct functional values are extracted from each of them. Thereby the total features become 352 ((16+16)*11). We have used open SMILE library [14] to extract features. We have differentiated one utterance from another by starting at the first non-0 pitch point and ends at the last non-0 pitch point.

The names of 16 low-level descriptors are:

- *Energy*
- *Mel-Frequency cepstral coefficients* (*mfcc*)(1-12)
- *Zero-crossing rate of time signal* (*frame-based*)
- *The voicing probability computed from the ACF.*
- *The fundamental frequency (F0)*

The functional values are maximum, minimum and their range, absolute position of maximum and minimum in frames, arithmetic mean, slope, offset, standard deviation, skewness and kurtosis.

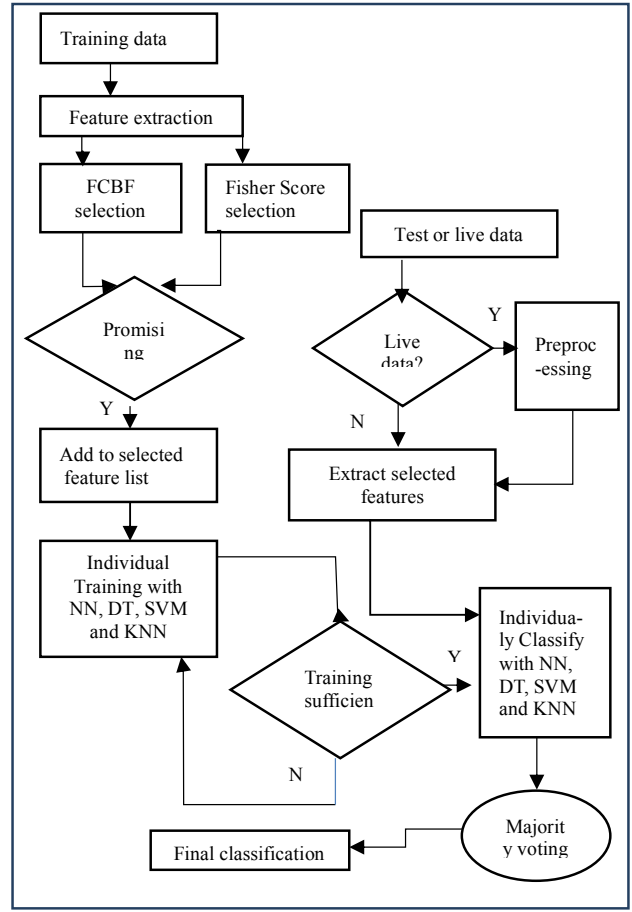

Fig. 1. Flow-chart of speech to emotion recognition

### 3.2 Feature selection

Emotion recognition from speech signal is nothing but a classification problem. To develop a system which can classify emotion correctly, system has to be trained sufficiently. Training largely depends on the quality of dataset. After selecting a dataset the main concern is to decide which features should be considered. A speech data has many cue, not all of them are equally responsible for emotion. As already told, we have extracted 352 features (feature set F) from speech signal.

If a feature set of 352 features is used to classify emotion it would be time consuming. Also there may exist redundant features and large feature set can cause curse of dimensionality [15].

Figure 2 shows a graph that represents a portion of 4 classes of emotional data taken from EMA dataset. Among all features only 12 features are plotted on its X-axis and their corresponding values are plotted on Y-axis. This figure shows that, irrelevant features exist on primary set F. For example for features 7 to 12, it has almost same value for all classes. So this feature is less informative to classify emotion. From this figure we can easily decide that feature only closely related to a particular class label and highly unrelated to other classes can help us to classify each emotion.

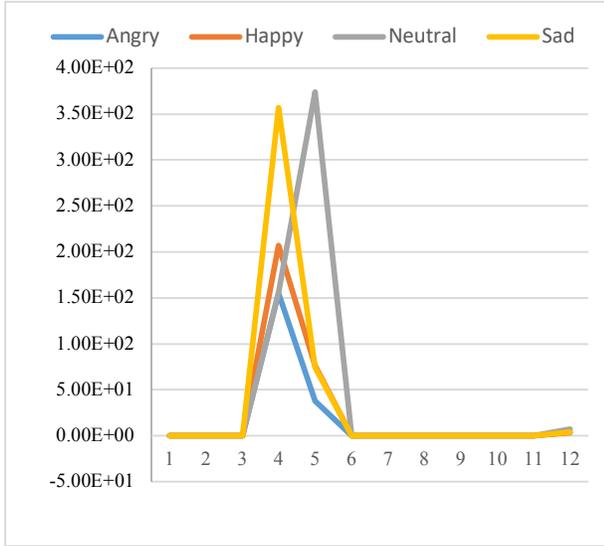

Fig. 2. Representation of 12 features before selection

We have applied two feature selection algorithms to choose the common features. The first one of them is FCBF. Here a feature is selected if and only if it satisfies the following two conditions: 1. Feature is highly correlated to a class and not to other classes and 2. Feature is not redundant [16].

FCBF uses symmetrical uncertainty (SU) to determine a feature either it is co-related to a class or not.

$$SU(X, Y) = 2 \left[ \frac{IG(X|Y)}{I(X)+I(X|Y)} \right] \quad (1)$$

For a feature x and class y, SU is calculated by Equation 1. where IG(.) denotes the information gain and I(.) denotes the entropy. After evaluating SU values only those features are selected in F′ feature set which have SU values larger than a threshold value. Then the features are sorted in decreasing order according to their SU values. And redundant features are removed from F′ [16].

The second chosen algorithm of feature selection is FS [17]. It calculates the Fisher Score of all features and we have selected those features having higher values. Equation for calculating Fisher Score of jth feature is:

$$F(X) = \frac{\sum_{i=1}^{c} n_i \left(u_i^j - u^j\right)^2}{\left(\sigma^j\right)^2}$$

$$\text{where} (\sigma^j)^2 = \sum_{i=1}^{c} n_i \left(\sigma_i^j\right)^2 \quad (2)$$

In Equation 2, $n_i$ is the size of the ith class, $\sigma_i^j$ is the standard deviation and $u_i^j$ is the mean of ith class corresponding to the jth feature. σj and uj are the standard deviation and mean for whole dataset corresponding to jth feature.

After combining those two, finally we have chosen 12 features to classify emotions. The representation of the selected features are shown in Figure 3. It shows that the selected features would help to detect classes effectively.

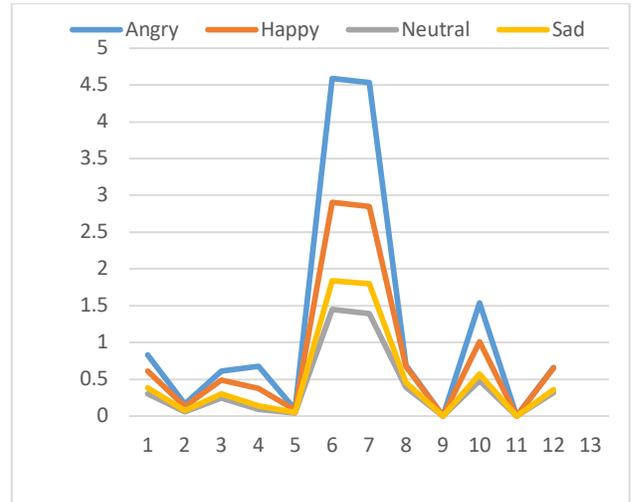

Fig. 3. Representation of selected 12 features.

### 3.3 Experimental Results

In order to recognize emotion, we have taken 280 examples from EMA dataset where each of four classes contains 70 examples. From Berlin dataset we have taken 339 examples where angry, happy, neutral and sad classes have 127, 71, 79 and 62 examples respectively. For training, 70% to 80% examples from both dataset have been used and the remaining examples are used for testing. Final result is calculated by averaging the result obtained from Berlin dataset and EMA dataset. A brief discussion of each classification algorithm is given below:

*1. Neural Network (NN):* We have used 3 layer Back propagation NN [16]. As there are 12 features and its need to detect 4 emotions, therefore the no. of nodes in the input and output layer of NN is 12 and 4 respectively. For the hidden layer 10 nodes are chosen. The highest accuracy is found for happy class which is 86.4% and the lowest accuracy is 69.23% and it is for neutral class and the average accuracy is 81.2%. The result is shown in Table 1.

TABLE 1
RESULT FOR NN.

| Class | Angry (%) | Happy (%) | Neutral (%) | Sad (%) |
|---|---|---|---|---|
| Angry | 84.62 | 0 | 15.38 | 0 |
| Happy | 0 | 86.4 | 13.6 | 0 |
| Neutral | 0 | 0 | 69.23 | 30.77 |
| Sad | 0 | 0 | 15.45 | 84.55 |

*2. K Nearest Neighbor (KNN):* This method estimates the local posterior probability of each class by the weighted average of class membership over the K nearest neighbors. We have run the algorithm for the no. of neighbor denoted as *k* from 4 to 15 and have noticed that the best result is found for *k* = 10. The result decreases for *k* < 10 and also for *k* > 10. The highest accuracy is 77.96% and it is for sad class whereas the lowest accuracy is 69.24% and it is for angry class and the average accuracy is 75%. Here among the four classifiers of our experiment, KNN performs worse. The result is shown in Table 2.

TABLE 2
RESULT FOR KNN

| Class | Angry (%) | Happy (%) | Neutral (%) | Sad (%) |
|---|---|---|---|---|
| Angry | 69.24 | 7.69 | 0 | 23.08 |
| Happy | 22.22 | 77.78 | 0 | 0 |
| Neutral | 0 | 0 | 75.02 | 24.98 |
| Sad | 0 | 0 | 22.04 | 77.96 |

*3. Decision Tree (DT):* In this experiment pruned DT is used. Here the highest accuracy is found for happy class which is 95.38% and the lowest accuracy is found for sad class which is 67.69%. The average accuracy of DT is 80.77% which is better than KNN. The result is shown in Table 3.

TABLE 3
RESULT FOR DT

| Class | Angry (%) | Happy (%) | Neutral (%) | Sad (%) |
|---|---|---|---|---|
| Angry | 80 | 20 | 0 | 0 |
| Happy | 0 | 95.38 | 4.62 | 0 |
| Neutral | 0 | 0 | 80 | 20 |
| Sad | 0 | 4.61 | 27.7 | 67.69 |

*4. Support vector machine (SVM):* Soft margin approach [17] has been used here to implement SVM. It chooses a hyperplane as cleanly as possible even if there is no hyperplane that can split the two classes. This method includes a variable $\xi_i$ which measures the degree of misclassification of data $X_i$ and it requires to solve the optimization problem of

$$\min_{w,\xi,b} \left\{ \frac{1}{2}||W||^2 + C \sum_{i=1}^{n} \xi_i \right\} \quad (3)$$

where the boundary of Equation 3 is $L_i(W.X_i - b) \geq 1 - \xi_i$ and $\xi_i \geq 0$ and $L_i$ is the label of $i^{th}$ class. It gives an average output of 81.69%. The result is shown in Table 4.

TABLE 4
RESULT FOR SVM

| Class | Angry (%) | Happy (%) | Neutral (%) | Sad (%) |
|---|---|---|---|---|
| Angry | 82.08 | 0 | 8.69 | 9.23 |
| Happy | 4.61 | 86.15 | 9.23 | 0 |
| Neutral | 0 | 0 | 78.53 | 21.47 |
| Sad | 10.77 | 0 | 9.24 | 80 |

*5. Majority Voting Technique:* NN, DT, KNN and SVM classifiers are executed in parallel. If at least two classifier's output is matched then the output class is decided as final emotion class, otherwise no decision is taken. Thus accuracy has increased up to 2-3% and average accuracy is 84.19% which is shown in Table 5.

TABLE 5
RESULT FOR MAJORITY VOTING TECHNIQUE

| Class | Accuracy (%) | Wrong Decision (%) | |
|---|---|---|---|
| | | Misclassification (%) | Can't Decide (%) |
| Angry | 84.06 | 14.46 | 1.5 |
| Happy | 88.15 | 11.45 | .5 |
| Neutral | 84.53 | 14.47 | 1 |
| Sad | 80.02 | 18.96 | 1 |

*6. Discussions:* Comparing the results obtained from four classification algorithms it can be pointed out that SVM gives more accurate result than others whereas KNN performs worst. KNN uses distance metric to perform classification, for better performance it needs to weight each feature according to correlation with correct class. But SVM can work with irrelevant feature as well. After SVM, NN shows better result than others and the result of training phase of NN is better than SVM. Another benefit of SVM model is that, it contains all the useful information, so classification does not need much time and also SVM is much more robust than NN. DT gives a result which is close to NN and much better than KNN. Although SVM generates better accuracy, majority voting technique has been used here to attain more accuracy.

One major problem of using majority voting technique is the adjustment of time. Unfortunately three different classifiers don't give result at the same time so it takes a little bit more time for final result. But when the importance of attaining better accuracy is higher than that of time requirement, majority voting technique performs better than any single classifier.

## 5. CONCLUSIONS

This paper recognizes emotion from human speech employing four machine learning techniques along with majority voting technique. The accuracy of emotion recognition from human speech highly depends on selected feature set i.e. if most relevant features can be selected, better accuracy can be achieved. Therefore most relevant and optimized feature set are selected by applying FCBF and FS algorithms. Then four machine learning techniques along with majority voting technique have been applied over those features of Berlin dataset and EMA dataset. From the experimental result it seems that because of feature set, all employed machine learning techniques have reduced the classification error. Moreover the use of majority voting technique has achieved a decent accuracy over other employed machine learning techniques. In case of emotion recognition from human speech as emotion from speech is not absolute i.e. same speech may represent multiple emotions; therefore the accuracy obtained in this simulation is satisfactory.